\documentstyle[aps,prl,twocolumn,epsf]{revtex}
\begin{document}

\title{Non-Universality in Random Matrix Ensembles with Soft Level
Confinement}
\author{C.~M.~Canali$^1$, Mats Wallin$^2$, and V.~E.~Kravtsov$^{1,3}$}
\address{$^1$International Centre for Theoretical Physics, 34100
Trieste, Italy\\
$^2$ Department of Theoretical Physics,
Royal Institute of Technology, S--100 44 Stockholm, Sweden\\
$^3$ Institute of Spectroscopy, Russian Academy of Sciences,
142092 Troitsk, Moscow r-n, Russia \\[3pt]
(Submitted to Phys.\ Rev.\ B, August 28, 1994)
%
\\ \medskip}\author{\small\parbox{14cm}{\small
Two families of strongly non-Gaussian random matrix ensembles (RME)
are considered. They are statistically equivalent to a one-dimensional
plasma of particles interacting logarithmically and
confined by the potential that has the long-range behavior
$V(\epsilon)\sim |\epsilon|^{\alpha}$ ($0<\alpha<1$), or
$V(\epsilon)\sim  \ln^{2}|\epsilon|$.
The direct Monte Carlo simulations on the effective plasma  model shows
that the spacing distribution function (SDF) in such RME can
deviate from that of the classical  Gaussian ensembles.
For power-law  potentials, this deviation  is
seen only near the origin $\epsilon\sim 0$, while for the
double-logarithmic potential the SDF shows the cross-over from the
Wigner-Dyson to Poisson behavior in the bulk of the spectrum.
\\[3pt]PACS numbers: 71.30.+h, 72.15.Rn, 05.60.+w}}\address{}\maketitle
\narrowtext

The classical theory of Random Matrices (RMT) developed by Wigner, Dyson
and Mehta \cite{mehta}
provides a statistical description of the energy levels in a variety of
quantum chaotic systems. In this way, one of  the simplest  statistical
characteristics is the probability distribution $P(s)$ of the spacing between
nearest-neighbor eigenvalues. In the framework of the classical RMT,
the spacing distribution function (SDF) follows very closely a universal
curve known as Wigner surmise \cite{mehta}. Its most important characteristic
is the vanishing of $P(s)$ at $s=0$, that demonstrates level repulsion.

In contrast, for classically non-chaotic systems, the random energy levels
are described by another universal distribution, the Poisson statistics,
which assumes all levels to be uncorrelated.

Both universal statistics are realized in a disordered system of
non-interacting electrons. The metal phase that exists for relatively
weak disorder, was proved \cite{Efetov} to be described by the Wigner-Dyson
statistics while the level statistics in the insulator phase is close to
the Poisson distribution.

The transition between this two phases known as
the Anderson transition, has much in common with the critical phenomena
in second order phase transitions and can be described by the scaling
approach \cite{Licciardello and Ramakrishnan}. Using scaling arguments, one
can show that in the critical region near the Anderson transition, there
should exist a third universal statistics \cite{Shklovskii}. The detailed
scaling analysis done recently \cite{rus. gang} showed this statistics
to be drastically
different from both the Wigner-Dyson and the Poisson statistics, the
corresponding spectral correlation functions being characterized by
nontrivial exponents related to the correlation length exponent $\nu$.

It is of great interest to see if a description of the critical
statistics in
terms of random matrices is still possible. Clearly, if this is the case,
the corresponding matrix ensembles must be of a completely different kind
from the ones belonging to the Wigner-Dyson universality class or from those
leading to the Poisson statistics. Our interest in this direction was
prompted by another recent discovery \cite {muttalib} of a new family of
random matrices, a
one-parameter solvable model, that displays a cross-over in the spacing
distribution from a highly correlated Wigner-Dyson to a completely
uncorrelated Poisson distribution when the parameter is varied.

In this paper we have tried to establish if the non-classical behavior of
level correlations is a generic feature shared a by broader class of random
matrix models. We will consider two strongly non-Gaussian ensembles of
random matrices. We
will show that the first one breaks the Wigner-Dyson universality only
locally, in the center of the spectrum. It will nevertheless allow us to
understand better the second class of models which is similar to
the exactly solvable model studied in Ref.\cite{muttalib}. Here we will show
that
this class of random matrices indeed breaks the  Wigner-Dyson universality
globally and displays a cross-over to a Poisson-like distribution.

Let us consider a physical system described by $N\times N$
random matrix $H$ whose eigenvalues $\{\epsilon_{n} \}, n=1,...N$ will
be also randomly distributed. Within the maximum entropy ansatz
\cite{Pichardreview,Beenacker}
for describing the eigenvalue distribution at a given mean level density
$\rho(\epsilon)$, we
can use the effective plasma model introduced by Dyson \cite{mehta},
where the joint probability density function $P(\{\epsilon \})$ is mapped
onto the Gibbs
distribution of a classical one-dimensional plasma of fictitious particles
with a pair-wise logarithmic repulsion $-\ln|\epsilon_{n}-\epsilon_{m}|$
and a one-particle potential $V(\epsilon)$ to keep the system confined:
\begin{equation}
\label{Gibbs}
P(\{\epsilon \})=Z^{-1}\,\exp[-\beta {\cal H}(\{ \epsilon_{n}\})],
\end{equation}

\begin{equation}
\label{H}
{\cal H}(\{ \epsilon_{n}\})= - \sum_{i<j} \ln|\epsilon_{i}-\epsilon_{j}|+
\sum_{i}V(\epsilon_{i}).
\end{equation}
Here $Z$ is the partition function and $\beta$, that plays the role of an
inverse temperature in the corresponding Gibbs ensemble, is related to
the symmetry of the original random matrix ensemble and is equal to 1,2
or 4 for orthogonal, unitary and symplectic ensembles, respectively
\cite{mehta}.

When the confining potential is quadratic, the model is the Gaussian
ensemble studied by Wigner, Dyson and Mehta and the corresponding SDF is
very close to the Wigner surmise. The inclusion of higher powers of
$\epsilon^{2}$ was shown \cite{Brezin Zee} to make no effect on the SDF in
the limit $N\rightarrow\infty$. Until very recently the confining
potential was believed to be irrelevant for level correlations in the
thermodynamic limit $N\rightarrow\infty$.

However, a recent work \cite{muttalib} has demonstrated that it is not the
case. For some specific one-parameter family of confining potentials
 $V_{p}(\epsilon)$, the exact solution in terms of the non-classical
$q$-orthogonal polynomials was found. It showed the level correlations to
deviate
 from the conventional Wigner-Dyson form as the parameter $p$
increases, the SDF
approaching the Poisson distribution for large values of $p$.
This was associated with the asymptotics of the confining potential
$V_{p}(\epsilon)\sim \ln^{2}|\epsilon|$ for $|\epsilon|\gg 1$
that is an extremely ``soft'' confinement as compared to the Gaussian confining
potential $V(\epsilon)=\epsilon^{2}$.

In this paper two main questions will be addressed. The first one is
how soft should the confining potential be in order to see deviations from
the classical Wigner surmise. The second question is
whether the exact solution found in Ref.\cite{muttalib} represents the
generic features of {\it all} models with the
double-logarithmic long-range behavior of the confining potential.

In order to answer the first of these questions, we consider a family of
power-law potentials:
\begin{equation}
\label{V}
V(\epsilon)=\frac{A}{2}|\epsilon|^{\alpha},
\end{equation}
where
$A>0$ and $\alpha>0$ are two constant parameters. For $\alpha=2$
Eq.(\ref{V})
reduces to the Gaussian quadratic confinement, while in the limit $\alpha
\rightarrow 0$ the combination of such power-law potentials reproduces the
double-logarithmic
potential $V(\epsilon)=\ln^{2}|\epsilon|=\lim_{\alpha\rightarrow
0}[\alpha^{-2}(|\epsilon|^{\alpha}-1)^{2}]$.

We will see that the level statistics exhibits {\it two} sharp transitions
when the parameter $\alpha$ decreases. The first one occurs at $\alpha=1$ and
it is connected with the break-down of translational invariance in the
eigenvalue space that is present for $\alpha\geq 1$ in the limit
$N\rightarrow\infty$. For $\alpha<1$, the SDF shows a non-classical,
$\alpha$-dependent behavior
 only near the center of the  spectrum $\epsilon=0$.
 The second critical value  is $\alpha=0$. For confining potentials that
increase only logarithmically, the SDF turns out to deviate from the
Wigner-Dyson form everywhere in the bulk of the spectrum.

The first transition is seen already within the mean-field approximation
suggested by Dyson \cite{mehta}. Let us define
$\rho(\epsilon)=\sum_{i}\delta(\epsilon-\epsilon_{i})$. By substituting
this definition into the Eq.(\ref{H}) one obtains the continuous version
of the energy functional ${\cal H}[\rho]$ in terms of $\rho(\epsilon)$.
The extremum of this functional corresponds to an equilibrium of the
effective plasma  expressed by the equation:

\begin{equation}
\label{MFE}
\int d\epsilon'
\langle\rho(\epsilon')\rangle\,\ln|\epsilon-\epsilon'|=V(\epsilon)+c,
\end{equation}
where $\langle\rho(\epsilon)\rangle$ is the mean density, and the Lagrange
multiple $c$ is to be found from the normalization
condition $\int \langle\rho(\epsilon)\rangle\, d\epsilon=N$.

Such a mean-field (MF) approximation
completely disregards the entropy part of the free-energy functional
and is exactly applicable only for $\beta=\infty$. However, the long-range
nature of the pair-wise interaction in Eq.(\ref{H}) makes the MF
approximation applicable in the bulk of the spectrum even at finite
$\beta$,
since for the class of confining potentials of Eq.(\ref{V}) the  relative
entropy contribution vanishes in the thermodynamic
limit as $(N \ln N)^{-1}$.

The solution $\rho_{\rm MF}(\epsilon)$ to the MF Eq.(\ref{MFE}),
confined to the region $-D<\epsilon<D$, can be found using the Cauchy
method \cite {Muskhelishvili} and is given by:
\begin{equation}
\label{MFS}
\rho_{\rm MF}(\epsilon)=\frac{1}{\pi^2}\sqrt{D^{2}-\epsilon^2}\, Re
\int_{0}^{D}\frac{dV/d\xi}{\sqrt{D^{2}-\xi^2}}\,\frac{\xi
d\xi}{\xi^2-\epsilon_{+}^2},
\end{equation}
where $\epsilon_{+}=\epsilon+i0$ and the band-edge $D$ is to be found from
the normalization condition.

For $\alpha\geq 1$ (strong confinement)
the main contribution to the integral in Eq.(\ref{MFS}) is made
 by the region $\xi\sim D$. In the thermodynamic limit $N\rightarrow\infty$
the band-edge is also divergent $D\rightarrow\infty$. Therefore, for any
fixed $|\epsilon|\ll D$, one can neglect the $\epsilon$-dependence in the
integrand of Eq.(\ref{MFS}). Then the mean level density tends to a constant
$\rho\sim N^{1-1/\alpha}$, signaling the translational invariance in the
$\epsilon$-space.

However, for $\alpha<1$ (weak confinement), the integral in Eq.(\ref{MFS})
is convergent even in the limit $D\rightarrow\infty$. The corresponding
limiting function
$\rho_{\rm MF}^{\infty}(\epsilon) \propto
|\epsilon|^{\alpha-1}$
can
be easily found as the limit
$z=\epsilon/D\rightarrow 0$
of the exact solution $\rho_{\rm MF}(\epsilon)$ to Eq.(\ref{MFS}):
\begin{equation}
\label{ex.sol}
\rho_{\rm MF}(\epsilon)=AC_{\alpha}\,
\frac{\sqrt{1-z^2}}{2\pi|\epsilon|^{1-\alpha}}\;
F\left(\frac{1}{2},\frac{1+\alpha}{2};\frac{3}{2};1-z^2
\right).
\end{equation}
Here $C_{\alpha}=\frac{\alpha^2
 2^{-\alpha}\Gamma(\alpha)}{\Gamma(\alpha/2)\Gamma(1+\alpha/2)}$,
$F(a,b;c;x)$ is a hypergeometric function, and the band-edge is given by
$D=2\left(\frac{N\Gamma^{2}(\alpha/2)}{2A\Gamma(\alpha)}
\right)^{1/\alpha}$.

Thus for $\alpha<1$ the mean  density  Eq.(\ref{ex.sol}) shows the
lack of translational
invariance in the large-$N$ limit and is singular at $\epsilon=0$.

This singularity, however, appears only in the MF approximation. An exact
treatment for $\beta=2$ that is based on the representation in terms of
orthogonal
polynomials,\cite{pap2} shows the value $\langle\rho(0)\rangle$ to be finite:
\begin{equation}
\label{r(0)}
\langle\rho(0) \rangle=\frac{(A/\pi)^{1/\alpha}}{(2/\alpha)\Gamma(1/\alpha)}
\sum_{i=0}^{\infty}\left[\frac{\Gamma(i+1/2)}{\Gamma(i+1)}
\right]^{2/\alpha}.
\end{equation}

Thus in case of weak confinement the MF approximation fails to describe the
mean level
density near the origin. It is natural to suppose that all the
level correlation functions will also have a non-classical form in
this region.

In order to study the correlation functions and in particular the SDF, we
have exploited the Coulomb plasma analogy and carried out systematic
Monte Carlo (MC) simulations on the one-dimensional classical system whose
probability distribution is given by Eqs.(\ref{Gibbs},\ref{H}).
As a check
that this method works and is numerically accurate we have first
studied
the three Gaussian ensembles whose density, two-point
correlation
functions and spacing distribution are exactly known \cite{mehta}. For
these systems MC turned out to
work extremely well for each of these quantities. For the power-law
potential Eq.(\ref{V})
for $\alpha <1$, we have carried out simulations of systems up to
$N=200$
particles. The simulations are very stable even for smaller N and we
have
typically worked with
$N=100$.
The evaluation of the
mean
density is straightforward. In Fig.~\ref{fig1} we plot this quantity for
$\alpha=0.5$ and $\beta=1,2,4$. The Monte Carlo result agrees very well
with
$\rho_{\rm MF}$ found from Eq.(\ref{ex.sol}) except around the origin, where
the
simulation is more accurate
and correctly gives a finite density at $\epsilon=0$. For $\beta=2$
the Monte Carlo value coincides
with that found from Eq.(\ref{r(0)}).
The simulations with
different numbers of particles illustrate another important property of
the
particle density for weak confinement ($\alpha <
1$), that is the ``incompressibility'' of the core of the particle-density
distribution.
In contrast to the $\alpha\geq 1$ case, for $\alpha <1$ the
confining potential is so weak that it does not ``compress'' particles
 in the core region near the origin.
 On adding more particles
to the system, these get positioned about the wings of the distribution,
rather than distribute themselves homogeneously throughout the spectrum,
as in the case of strong confinement ($\alpha\geq 1$). The particle density
$\rho(\epsilon)$ in the
core region  is almost independent of the number of
particles but depends on the inverse temperature $\beta$.

The latter dependence is also a characteristic feature of the weak
confinement. For strong confinement, the $\beta$-dependence is present only
in $1/N$ corrections to the mean density and thus is negligible. It leads,
in particular, to the independence of the mean level density of the
symmetry of the Hamiltonian. For random matrix ensembles with weak
confinement considered here, all the $\beta$-dependence is ``accumulated'' in
the
core region near the origin that contains a few levels on the average.

The MC evaluation of the SDF is, in principle, also straightforward.
However, in order to compare it with the Wigner surmise
we need to rescale the particle positions $\epsilon$ so that
the average spacing between two adjacent ones is one. This is known as an
``unfolding procedure'' and is always used in numerical calculations of
spectral correlations \cite{KramerOhtsuki}. It consists in introducing the
new variable $\sigma$ instead of $\epsilon$ according to a map:
\begin{equation}
\label{sigma}
\sigma(\epsilon)=\int_{0}^{\epsilon}\langle\rho(\epsilon')\rangle
\,d\epsilon'.
\end{equation}
The mean density is trivially unity as a function of this variable.

In order to study the SDF in the {\it bulk}, we use the MF solution
Eq.(\ref{ex.sol}) for unfolding according to Eq.(\ref{sigma}).
We checked that the obtained unfolded mean density is consistently equal
to one, except close to the origin and the band-edge.
The unfolded spacing
$\tilde P(\sigma)$ turned out, within our numerical
accuracy, identical to the Wigner surmise for any $\alpha$. Therefore,
 in the {\it bulk} of the spectrum, the Wigner-Dyson universality holds for
the power-law weakly confining potentials.

However this universality is broken around the origin. To show this
we
consider a reference particle fixed at the origin. The unfolded spacing
must
be evaluated in a different way here, since the MF density is not
accurate.
Therefore we perform the unfolding by
computing
\begin{equation}
\tilde P(\sigma) = \biggl[{P(\epsilon)\over \rho(\epsilon)}\biggr]_{\epsilon =
\epsilon(\sigma)}\;,
\end{equation}
where the function $\epsilon(\sigma)$ is obtained by inverting
numerically
Eq. (\ref{sigma}) and using for $\rho(\sigma)$ the density evaluated
by
MC simulations. The result is shown in Fig.~\ref{fig2}, where we plot the
unfolded
SDF for few values of $\alpha<1$ and $\beta=1$ in comparison with the
classical spacing of the Gaussian ensemble.
We can clearly see that $\tilde P(\sigma)$ for small $\sigma$
 does not follow the Wigner-Dyson universal
 behavior $\sigma^{\beta}$ and starts out roughly like
$\sigma^{\beta/\alpha}$.
If we assume that the new
variable
$\sigma$ is proportional to $\epsilon^{\alpha}$, as obtained from
Eq.(\ref{sigma}) using $\rho(\epsilon)\propto \epsilon^{\alpha-1}$,
this behavior would correspond to
$P(\epsilon)/\rho(\epsilon)\propto\epsilon^{\beta}$.
Notice also that the decay of the SDF for $s\gg 1$ depends on
$\alpha$
and is slower than that for the Wigner-Dyson distribution. We conclude that
for the
power-law weak confining potential, the
Wigner-Dyson universality is broken only locally around $\epsilon\sim 0$.
This conclusion is also reached for $\beta=2$, using the independent method
of orthogonal polynomials. \cite{pap2}

Now we consider the second class of random matrices, with the confining
potential that behaves asymptotically like $V(\epsilon)\propto
\ln^{2}|\epsilon|$. Since our goal is to study the eigenvalue correlations
{\it in the bulk} of the spectrum, we choose for numerical simulations the
regularized confining potential that is equal to zero at the origin:
\begin{equation}
\label{double}
V(\epsilon)=\frac{A}{2}\ln^{2}(1+B|\epsilon|),
\end{equation}
where $A$ and $B$ are parameters of order 1.

In Fig.~\ref{fig3} we show the {\it bulk} SDF for $A=1, 0.5, 0.2, 0.1$,
for the orthogonal symmetry ($\beta=1$), together
with the spacing distribution of the corresponding
Gaussian ensemble for comparison.
We can clearly see that for small enough $A$ the spacing distribution
departs from the Wigner  distribution and shows an
incipient tendency to become more Poisson-like when A is further reduced.
Similar deviations from the Gaussian ensemble occur also for
the unitary and symplectic case ($\beta=2,4$).

This is very similar to the cross-over found analytically in Ref.
\cite{muttalib} for the exactly solvable model with the
double-logarithmic long-range behavior of the confining potential.
We can conclude, therefore, that the cross-over is indeed
not an exclusive property of the exactly solvable model and is more likely
a generic feature
shared by all the random matrix ensembles with the double-logarithmic
asymptotics of the confining potential.

The cross-over in the spacing distribution displayed by this family of
random matrices is remarkably similar
to the transition observed in exact numerical
calculations \cite{Shklovskii,muttalib,num} on finite
disordered systems going through the Anderson transition.

\acknowledgments %
We would like to thank K.~A.~Muttalib and E.~Tosatti for discussions. We
also grateful to M.~J.~P.~ Nijmeijer for helping us to get started with the
Monte Carlo simulations and for useful comments throughout this project.
M.W.\ is supported by the Swedish Natural Science Research Council.

\begin{figure}
\caption{Particle density for power-law potential with $\alpha=0.5$:
The Monte Carlo results for $\beta=1,2,4$ are plotted vs. the MF density.}
\label{fig1}
\end{figure}

\begin{figure}
\caption{Nearest-neighbor spacing distribution in the middle of the spectrum
for $\beta=1$ and different values of $\alpha$. The $\alpha=2$ case corresponds
to the Gaussian Orthogonal Ensemble.}
\label{fig2}
\end{figure}

\begin{figure}
\caption{Nearest-neighbor spacing distribution
for the logarithmic confining potential, measured in the bulk of
the spectrum at $\beta=1$. By decreasing the parameter $A$, the
spacing deviates from the
universal Wigner-Dyson distribution approaching the Poisson distribution
(both also plotted).}
\label{fig3}
\end{figure}
\end{document}